\documentclass[]{mn2e}
\usepackage{graphics}
\usepackage{epsf}
\usepackage{subfigure}

\begin{document}


\title{Principal Component Analysis of RR Lyrae light curves}
\author[Kanbur and Mariani]{S. M. Kanbur$^{1}$\thanks{Email: shashi@astro.umass.edu} and H. Mariani$^{1}$ 
\\
$^{1}$Department of Astronomy, University of Massachusetts
\\
Amherst, MA 01003, USA}


\date{Received XX XXX 2003 / Accepted XX XXX 2003}

\maketitle

\begin{abstract}
In this paper, we analyze the structure of RRab star light curves
using Principal
Component Analysis. We find this is a very efficient way to describe many
aspects of RRab light
curve structure: in many cases, a Principal Component fit with
9 parameters can describe a RRab light curve including bumps whereas a 17
parameter Fourier fit is needed. As a consequence we show
show statistically why the amplitude is also a good
summary of the structure of these RR Lyrae light curves. We also use our analysis
to derive an empirical relation relating absolute magnitude to light curve
structure. In comparing this formula to those derived from
exactly the same dataset but using Fourier parameters,
we find that the Principal Component Analysis approach has distinct advantages. 
These advantages are, firstly, that the errors on the
coefficients multiplying the fitted parameters 
in such formulae are much smaller, and secondly,
that the correlation
between the Principal Components is significantly smaller than the correlation
between Fourier amplitudes. These two factors lead to reduced formal errors, in some
cases
estimated to 
be a factor of 2,
on the eventual fitted value of the absolute magnitude. This technique will
prove very useful in
the analysis of data from existing and new large scale survey projects
concerning variable stars.

\end{abstract}



\section{Introduction}

Kanbur et al (2002), Hendry et al  (1999), Tanvir et al (2004) introduced the use of
Principal Component Analysis (PCA) in studying Cepheid light curves.
They showed
that a major advantage of such an approach over the traditional
Fourier method is
that it is much more efficient: an adequate Fourier description
requires, at best, a fourth order fit or
9 parameters, whilst a PCA analysis requires only 3 or 4 parameters
with as much as 81$\%$ of the
variation in light curve structure being explained by the first parameter.
Later, Leonard et al (2003)
used the PCA approach to create Cepheid light curve templates to estimate
periods and mean
magnitudes for HST observed Cepheids. The purpose of this
paper is to apply the PCA technique to the study of RR Lyrae light curves.

The mathematical formulation and error characteristics of PCA are given
in K02 and will only be
summarized here.

\section{Data}

The data used in this study were kindly supplied by Kovacs
(2002 private communication)
and used in Kovacs and Walker (2001, hereafter KW).
These data consist of 383 RRab stars with well observed V band light curves
in 20 different globular clusters. KW performed a Fourier fit to these
data, which, in some cases, is of order 15. Details concerning the data can be
found in KW.
The data we work with in this paper is this Fourier fit to the magnitudes
and we assume that the
Fourier parameters published by KW are an accurate fit to the actual
light curves.
We start with the data in the form used in KW: a list of the mean magnitude,
period and Fourier
parameters for the V band light curve. The light curve can thus be
reconstructed using an expression of the
form
\begin{eqnarray}
V = A_0 + \sum_{k=1}^{k=N}{A_ksin(k\omega t + {\phi}_k)},
\end{eqnarray}
where $A_0$ is the mean magnitude, ${\omega}=2\pi/P$, $P$ the period, $A_k,{\phi}_k$
the Fourier parameters given in KW. These light curves are then rephased so
that maximum light occurs at
phase 0 and then rewritten as
\begin{eqnarray}
V = A_0 + \sum_{k=1}^{k=N}(a_k cos(k\omega t) + b_k sin(k\omega t)).
\end{eqnarray}
The $a_k,b_k$ are the light curve characteristics entering into the PCA
analysis (K02).
We then solve equation (4) of K02, either after, or before removing an average
term from the Fourier
coefficients in equation (2). With PCA, the light curve is written as a sum of
"elementary" light curves,
\begin{eqnarray}
V(t) = PCA1.L_1(t) + PCA2.L_2(t) + PCA3.L_3(t)+.....,
\end{eqnarray}
where $V(t)$ is the magnitude at time t, $PCA1, PCA2..$ etc. are the PCA
coefficients and the
$L_i(t), i=1,2,3...$ are the elementary light curves at phase or time t.
These elementary light curves are not a priori given, but are estimated from the 
dataset in question.
Each star
has associated with it a set of coefficients $PCA1, PCA2,...$ and these can be
plotted
against period just as the Fourier parameters in equation (1) are plotted
against
period. We also note that the PCA results are achieved as a result of
the analysis
of the {\it entire} dataset of 383 stars whereas the Fourier method produces
results for stars individually. This feature of PCA is particularly useful when performing an
ensemble analysis of large numbers of stars obtained from projects such as
OGLE, MACHO and GAIA.

\section{Results}

Solving equation (4) of K02 yields the Principal Component scores and the
amount of variation
carried by each component. What we mean by this is the following: if we carry
out an $N^{th}$ order PCA fit, then PCA will assume that all the
variation in the dataset is described by $N$ components and simply
scale the variation carried by each component accordingly.
Table 1 shows this "amount of variation" quantity with and without
the average term removed. We see that in the case when
we do not remove the average term the first PC explains as much as
$97\%$ of the variation in the light curve structure.
In the case when we do remove the average term from the Fourier coefficients,
the first PCA
coefficient 
explains as much 81 percent of the variation in light curve structure. In
either case, the first four
components explain more than $99.99\%$ of the variation.

Figures 1 and 2 show some representative light curves from our RRab dataset.
In each panel of these two figures, the solid line is the Fourier decomposition of order 15 
(that is 31 parameters) used by KW, whilst the dashed line is a PCA generated light curve of
order 14 (that is 15 parameters).
Straightforward light curves such
as the one given in the bottom and top left panels of figures 1 and 2 respectively are easily reproduced
by our method. The top left panel of figure 1 provides an example of an RRab light curve
with a dip and sharp rise at a phase around 0.8. This is well reproduced by PCA. It could be argued
that PCA does not do as well as Fourier in mimicking this feature, for example, in the bottom right panel
of figure 2. However, the difference in the peak magnitudes at a phase of around 0.8 is of the order of 0.02mags. It is also important to remember that
the PCA method is an ensemble method and analyzes all stars in a dataset
simultaneously. With Fourier, it is possible to tailor a decomposition
to one particular star. This difference can be seen either as a positive
or negative point about either technique. Given this, we contend that PCA
does remarkably well in describing the full light curve morphology of
RRab stars.
On the other hand, the Fourier curve in the bottom left panel of figure 2 at this phase is not as smooth as the
PCA curve.

In fact the PCA curves do not change much after about 8 PCA parameters. Even
though table 1 implies that the higher order PCA eigenvalues are small,
we feel justified in carrying out such a high order PCA fit because its
only after about 8 PCA components that the fitted light curve assumes
a stable shape. The left panel  of figure 3 displays
an eighth order PCA fit (9 parameters, dashed line) and a fourth order Fourier fit (9 parameters, solid
line). The Fourier curve still has some numerical wiggles whilst the
PCA curve is smoother. In addition, the
two curves disagree at maximum light. The right panel of figure 3 shows, for the same star, the same order
PCA curve as the left panel and an eighth order Fourier fit (17 parameters). Now the two light curves agree very well.
Note that in portraying the PCA and Fourier fits of reduced order in this figure, we simply truncated the original
representations to the required level. 

We suggest that figures 1-3 and table 1
provide strong evidence that PCA is an {\it efficient} way to describe RRab
light curve structure without compromising on what light curve features are
captured by this description. 

Figures 4-6 display plots of the first three PC scores plotted against log
period
for our sample. The errors associated with these PCA scores are discussed in
section 4 of K02 and given in
equation 6 of that section. The orthogonal nature of these scores may well
provide insight into the physical processes causing observable features in the light curve structure. A detailed study of these plots, in conjunction with
theoretical models, is left for a
future paper.

Figure 7 graphs V band amplitude against the first PCA coefficient
(after averaging).
We see a very tight correlation. Since table 1 implies that PCA1 explains about $81\%$ of the variation in
light curve structure, figure 6 shows that the amplitude is a good
descriptor of RRab light curve
shape, at least for the data considered in this paper. Although the Fourier amplitudes are also
correlated with amplitude, with PCA, we can quantify, very easily, the amount of variation described by each PCA component.
This has implications for both modeling and observation. On the modeling side, a computer code that
can reproduce the observed amplitude at the correct period, will also do a good job of reproducing the
light curve structure. On the observational side, this provides insight into
why we can use the amplitude, rather than a full blown PCA
or Fourier analysis,
to study the {\it general} trends of light curve structure.
This is why comparing
theoretical and observational RRab light curves
on period-amplitude diagrams works reasonably well, though we caution that
a careful analysis should consider the finer
details of light curve structure.  

Figures 6 and 7 display plots of the first two PCA coefficients and Fourier
amplitudes, respectively, for our data,
plotted against each other. Whilst $A_1$ and $A_2$ are correlated with each
other, $PCA1$ and
$PCA2$ are not, by construction. A similar situation
would occur had we plotted $A_1$ or $A_2$ against $A_3$. This is another
advantage of PCA analysis of variable star light curves:
the different PCA components are orthogonal to each other. A practical
advantage of this feature is outlined in the
next section. 

\begin{figure*}
        \vspace{0cm}
        \hbox{\hspace{0.2cm}\epsfxsize=7.5cm \epsfbox{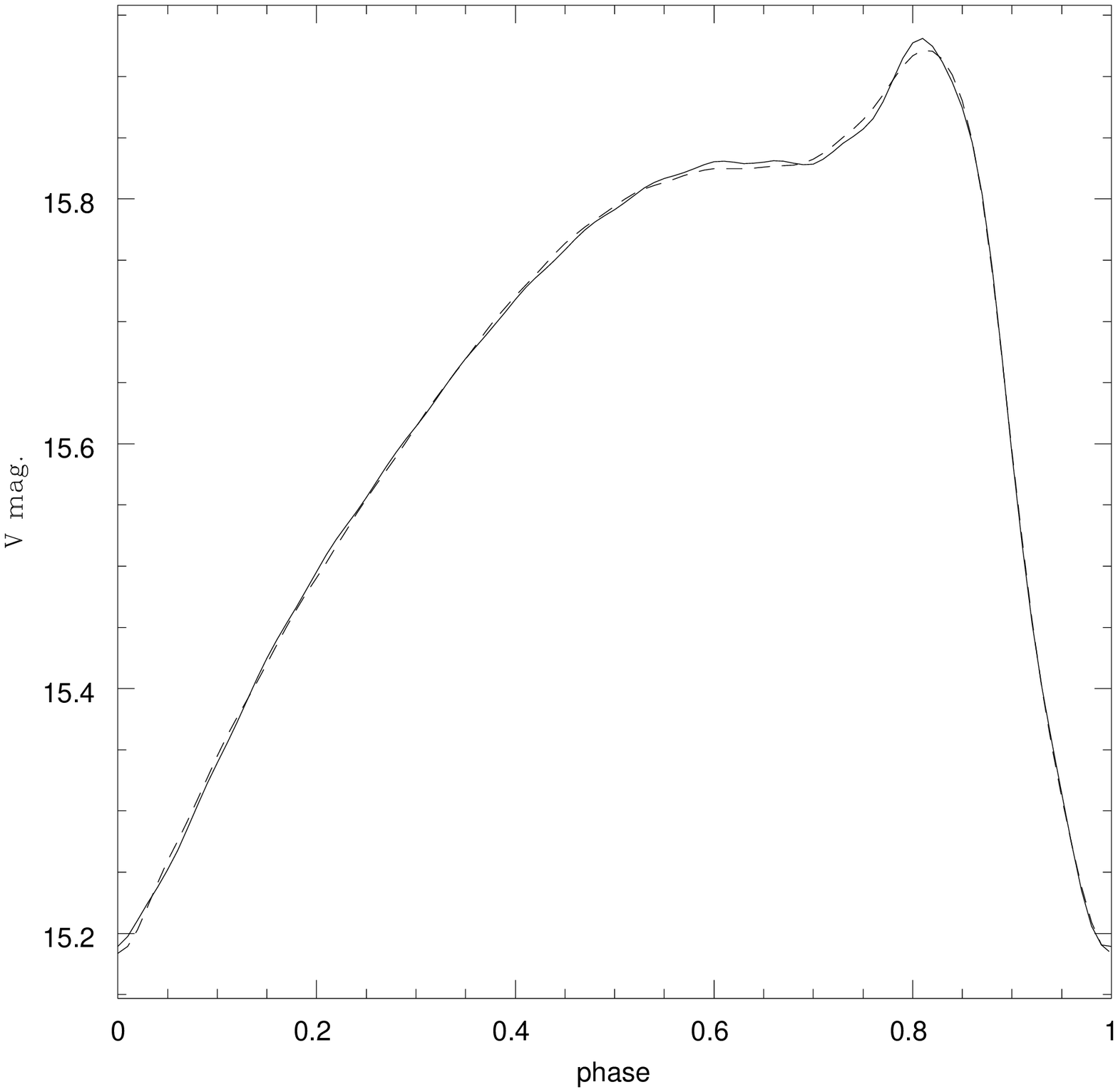}
        \epsfxsize=7.5cm \epsfbox{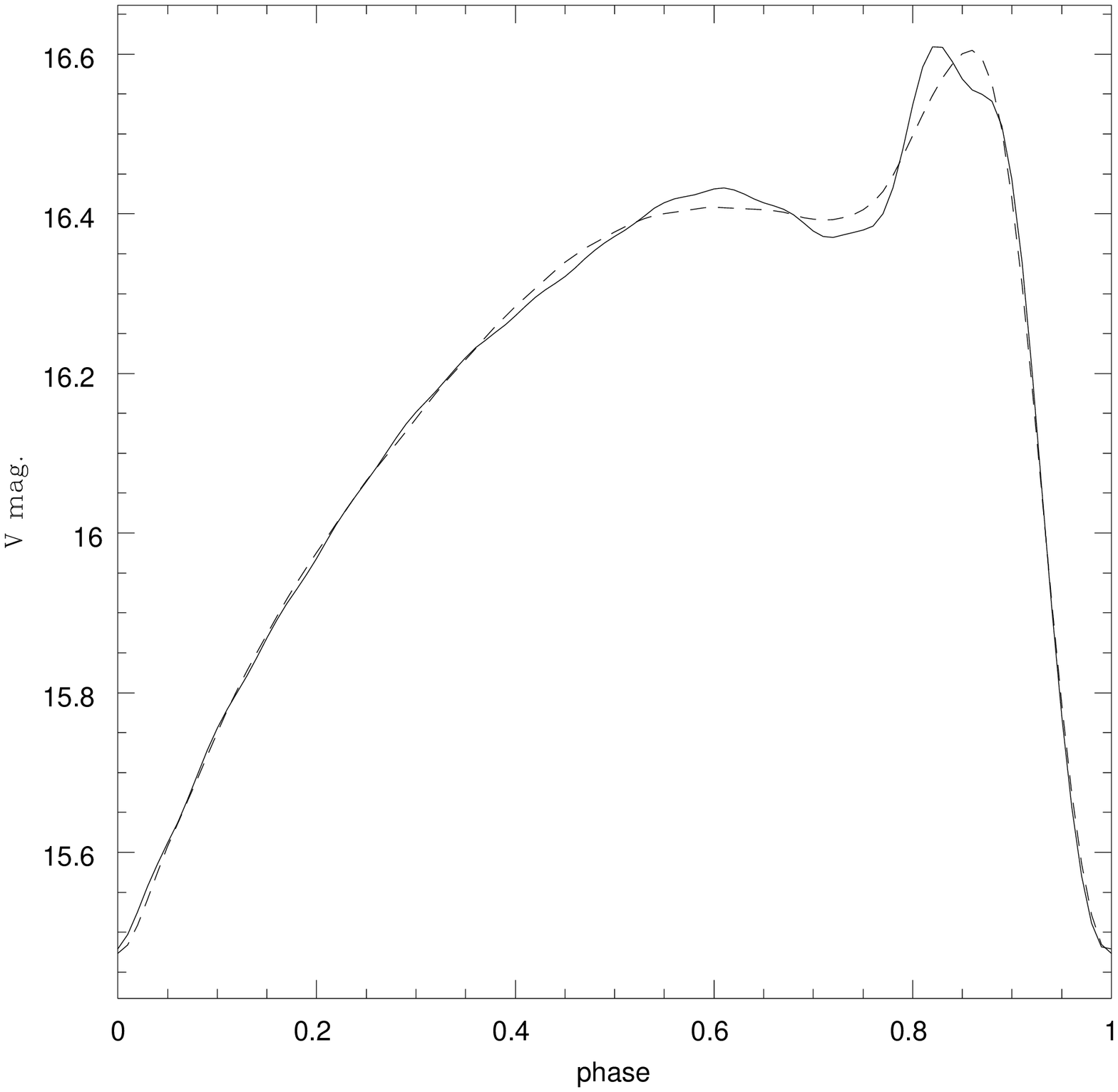}}
        \hbox{\hspace{0.2cm}\epsfxsize=7.5cm \epsfbox{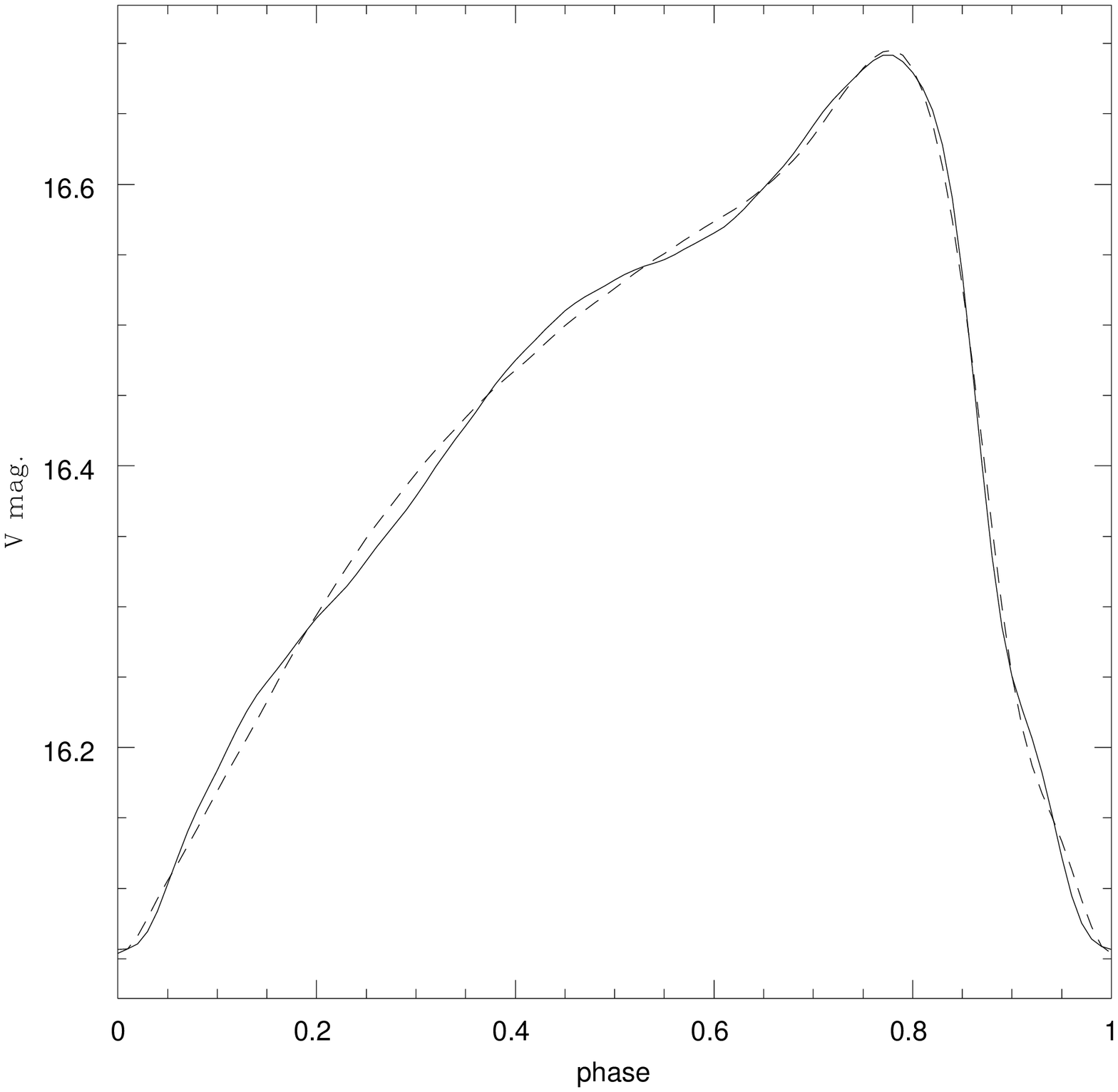}
        \epsfxsize=7.5cm \epsfbox{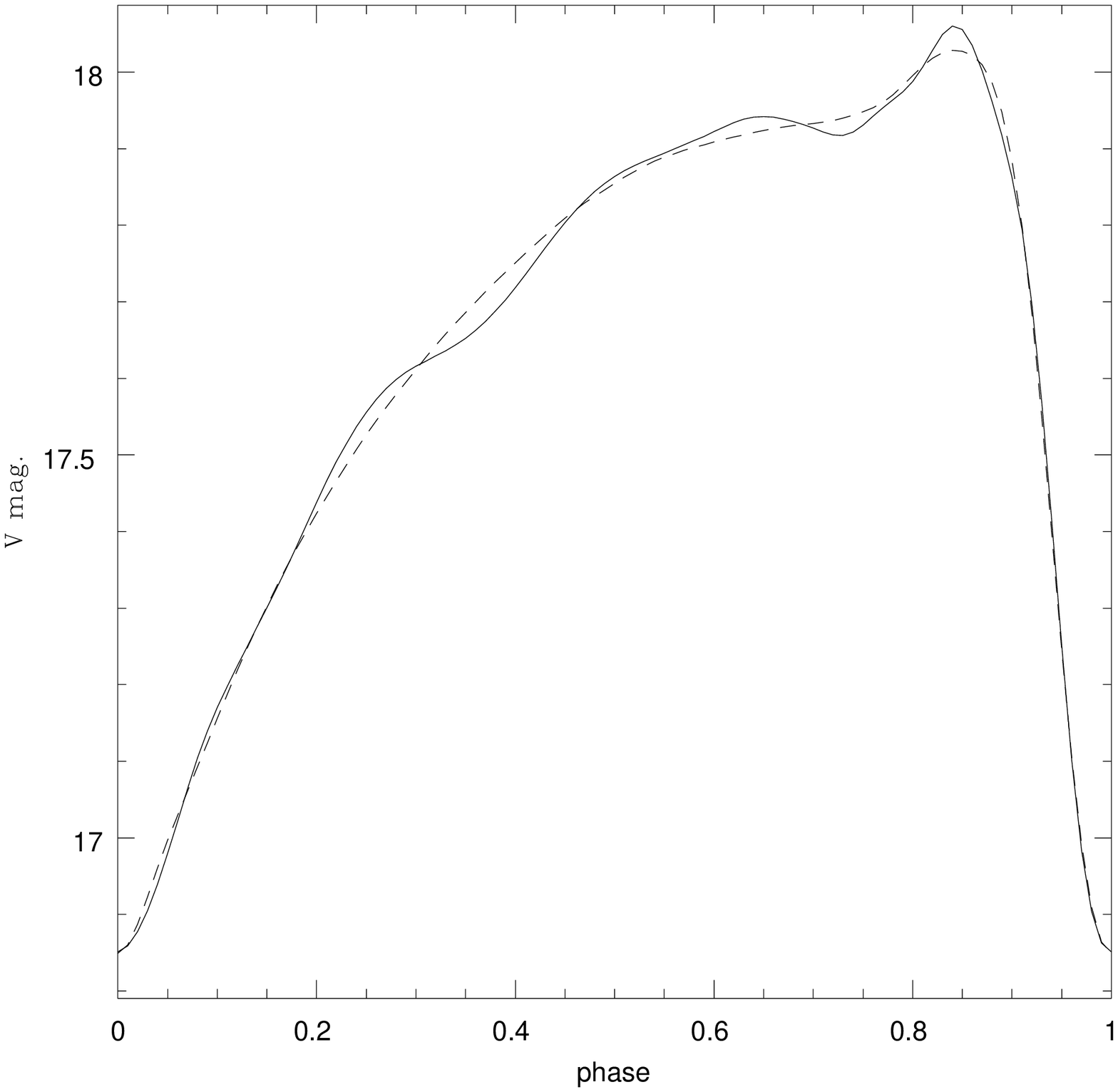}}
        \vspace{0cm}
        \caption{Light curve reproduction using Fourier (solid lines) and PCA (dashed lines) methods \label{fig1}}
        \end{figure*}

\begin{figure*}
        \vspace{0cm}
        \hbox{\hspace{0.2cm}\epsfxsize=7.5cm \epsfbox{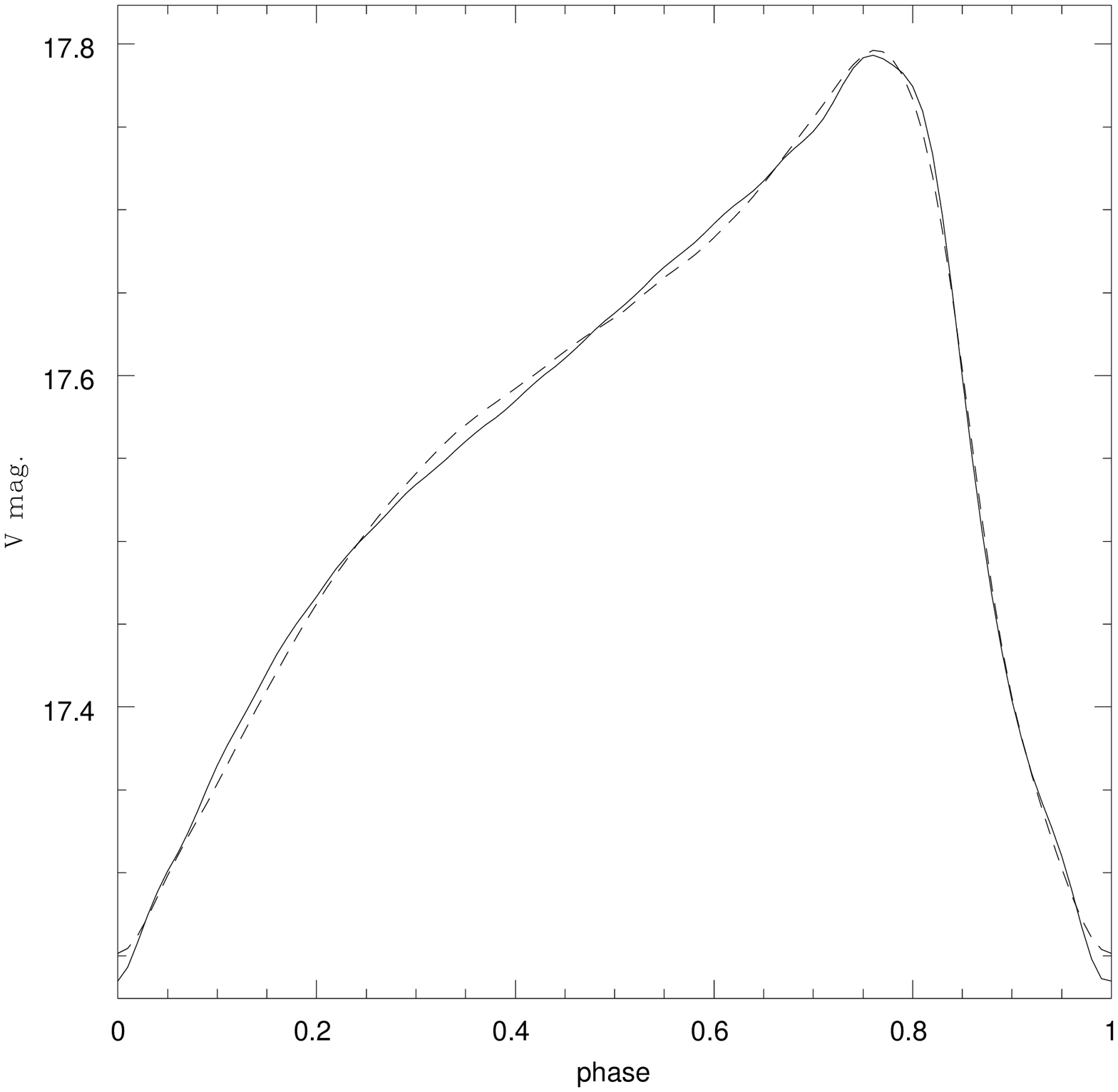}
        \epsfxsize=7.5cm \epsfbox{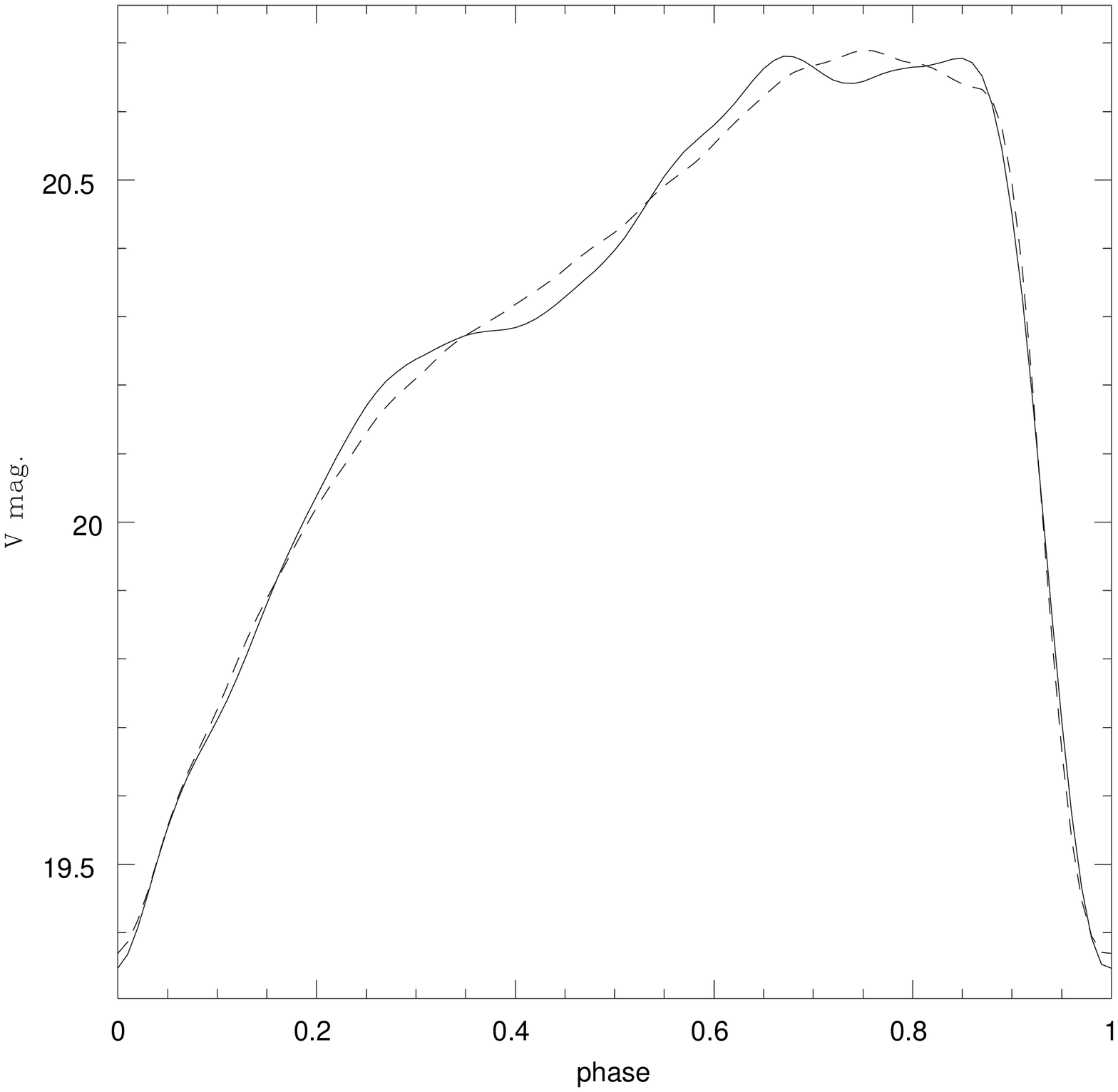}}
        \hbox{\hspace{0.2cm}\epsfxsize=7.5cm \epsfbox{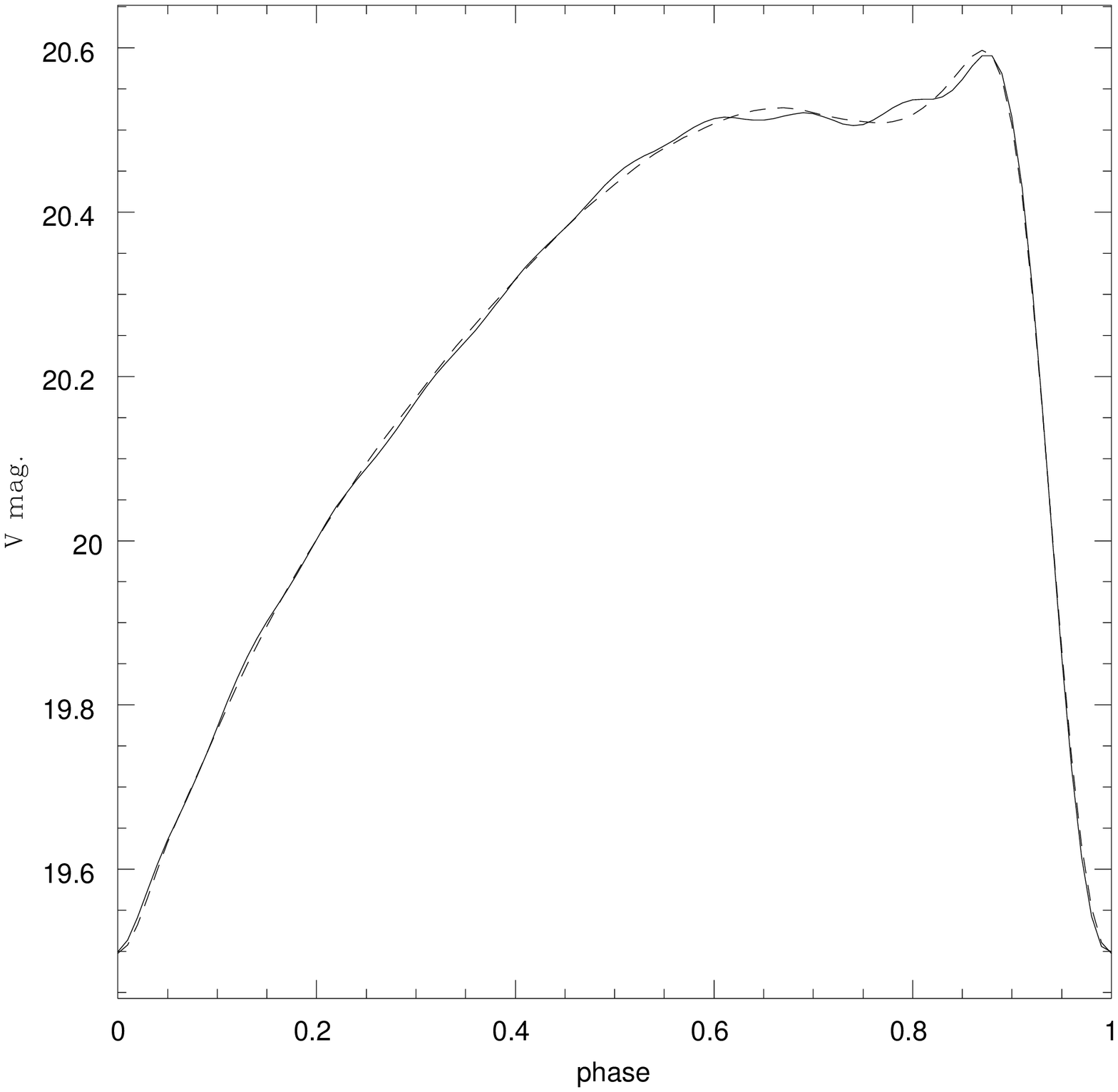}
        \epsfxsize=7.5cm \epsfbox{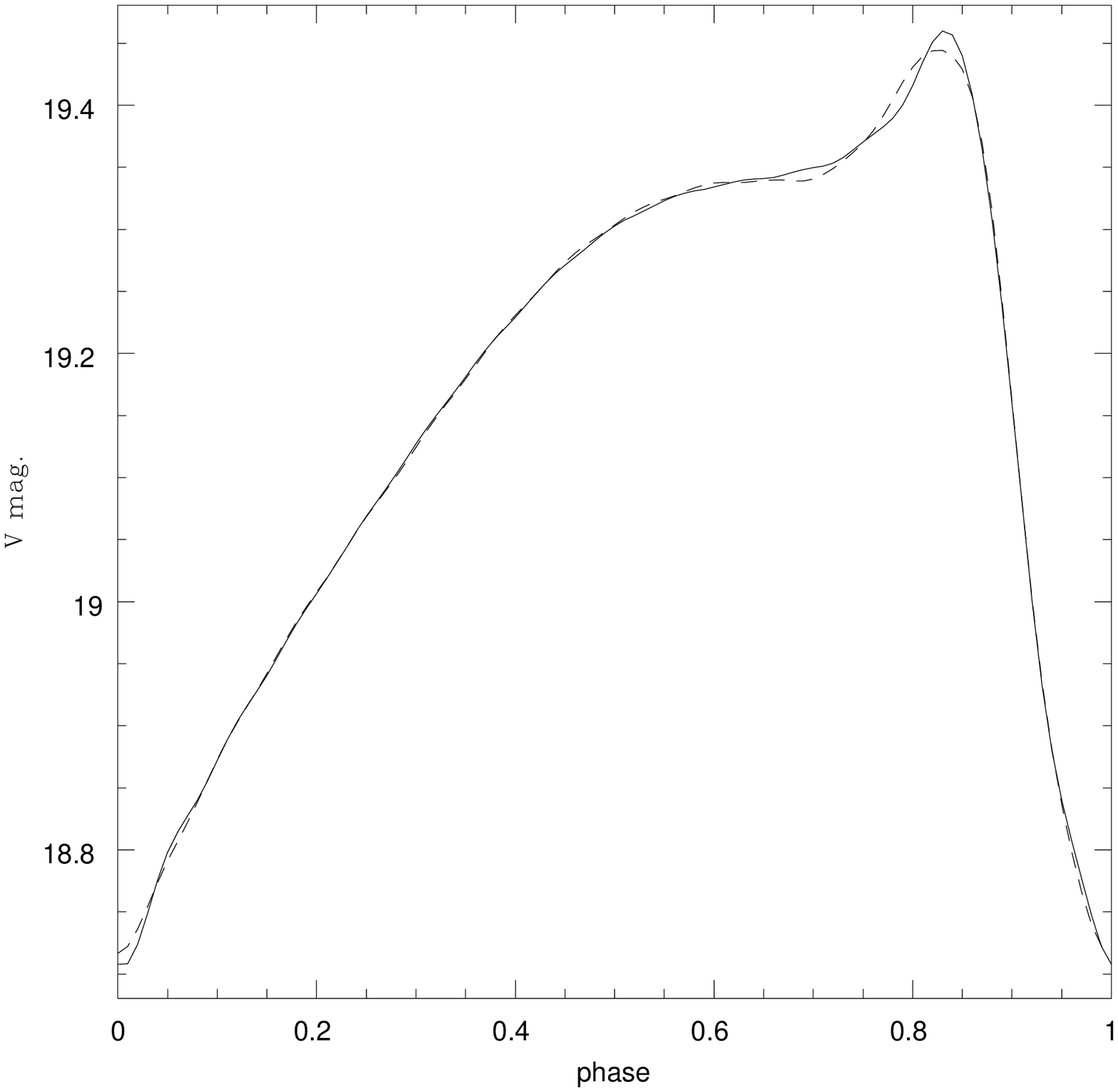}}
        \vspace{0cm}
        \caption{Light curve reproduction using Fourier (solid lines) and PCA (dashed lines) methods \label{fig1}}
        \end{figure*}
\begin{figure*}
        \vspace{0cm}
        \hbox{\hspace{0.2cm}\epsfxsize=7.5cm \epsfbox{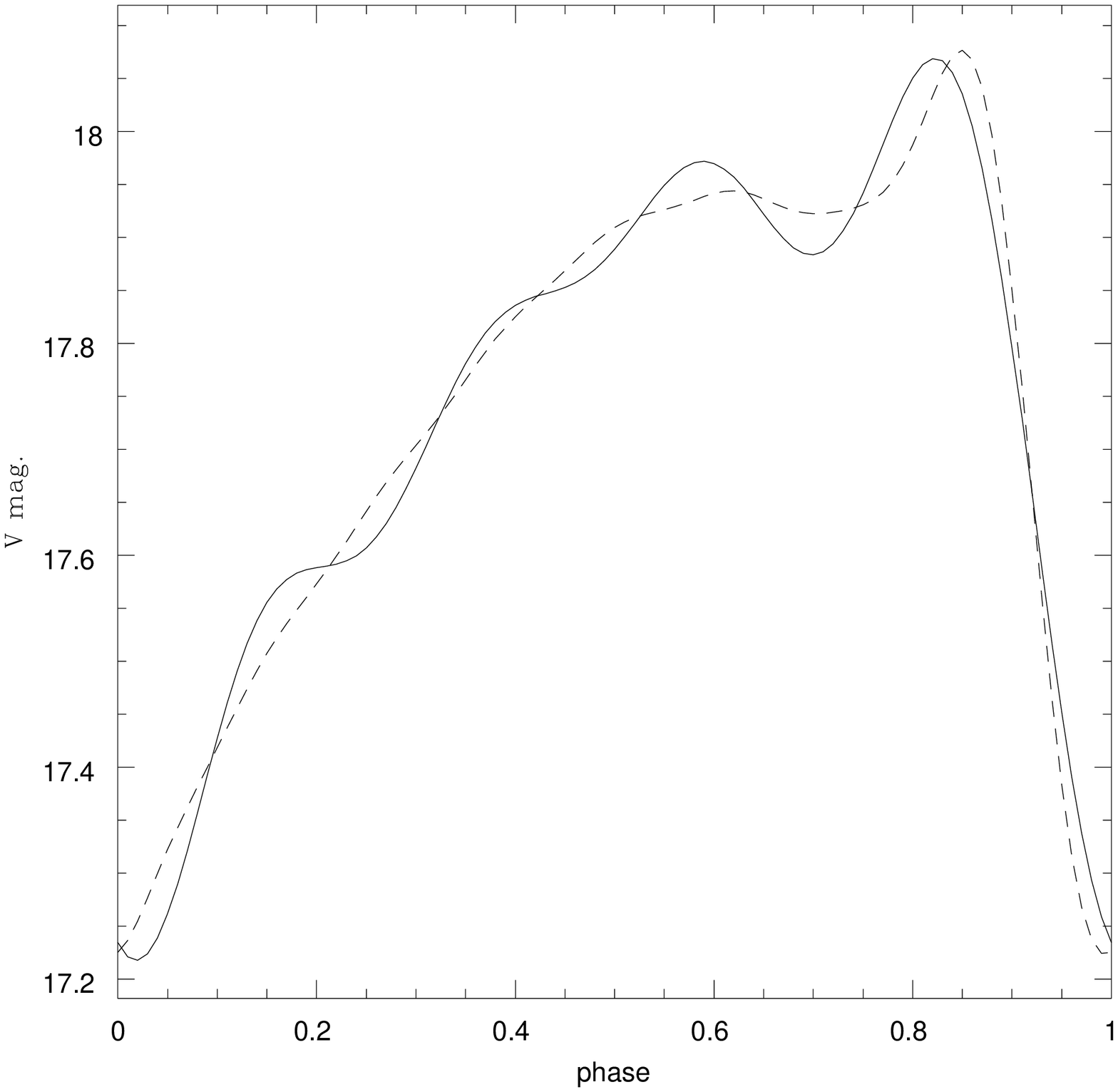}
        \epsfxsize=7.5cm \epsfbox{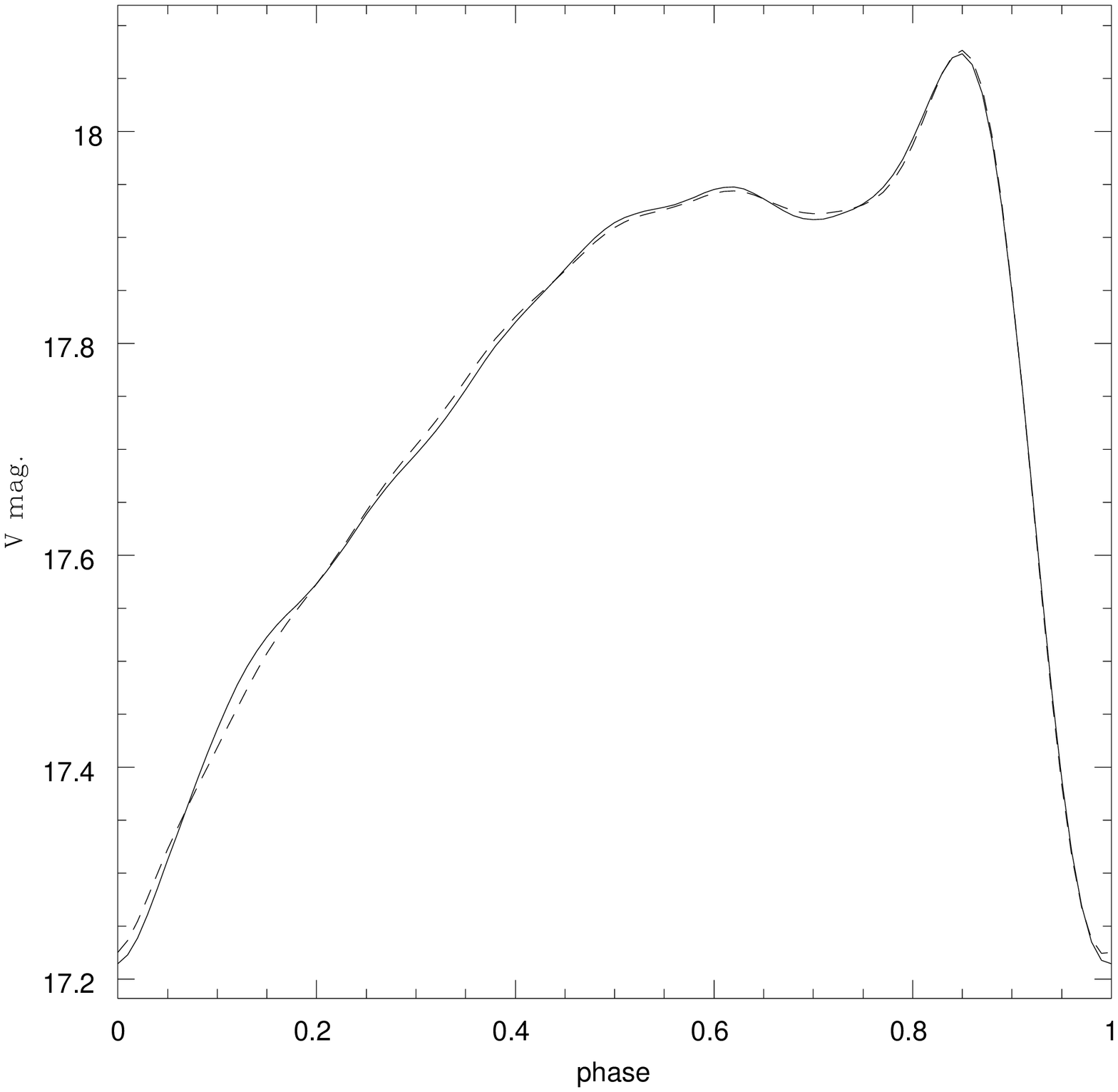}}
        \vspace{0cm}
        \caption{Light curve reproduction using Fourier (solid lines) and PCA (dashed lines) methods. The left panel is a fourth order (9 parameters) Fourier
fit and an eight order PCA (9 parameters) fit. The right panel is an eight order (17 parameters) Fourier fit and an eight order
PCA (9 parameters) fit. ( \label{fig1}}
        \end{figure*}

\begin{figure}
\hbox{\hspace{0.1cm}\epsfxsize=8.0cm \epsfbox{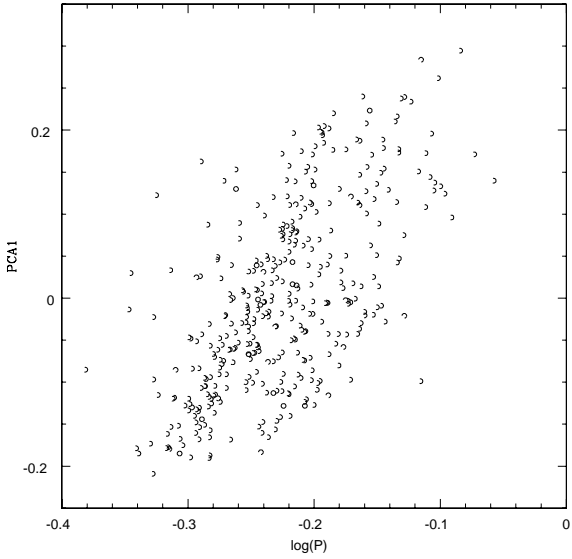}}
\caption{Plot of first Principal Component against log period.}
\label{ref}
\end{figure}

\begin{figure}
\hbox{\hspace{0.1cm}\epsfxsize=8.0cm \epsfbox{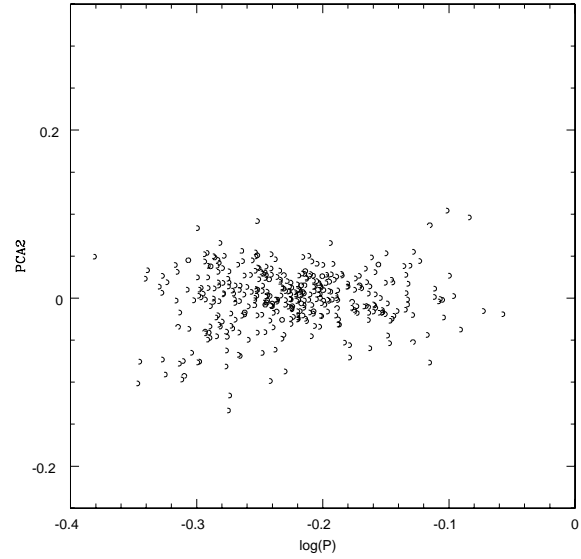}}
\caption{Plot of second Principal Component against log period.}
\label{ref}
\end{figure}

\begin{figure}
\hbox{\hspace{0.1cm}\epsfxsize=8.0cm \epsfbox{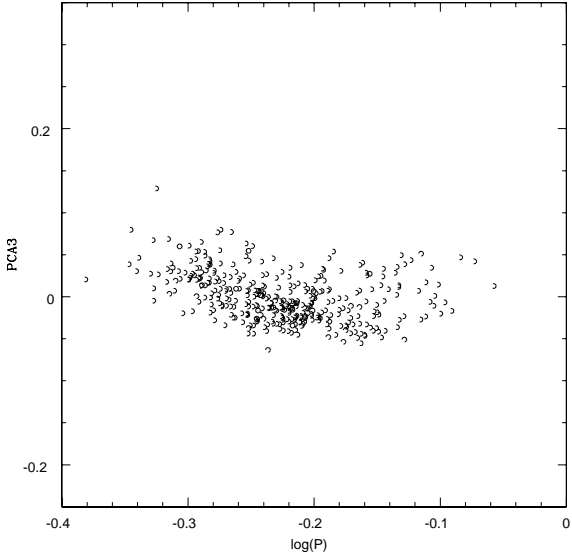}}
\caption{Plot of third Principal Component against log period.}
\label{ref}
\end{figure}

\begin{figure}
\hbox{\hspace{0.1cm}\epsfxsize=8.0cm \epsfbox{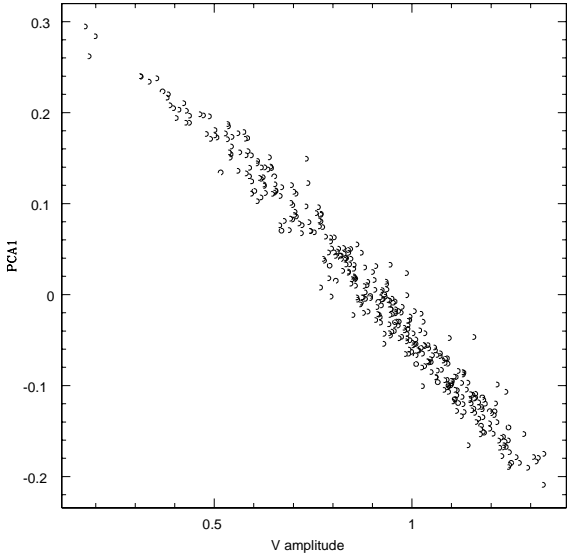}}
\caption{ Plot of V band amplitude against the first PCA coefficient.}
\label{ref}
\end{figure}

\begin{figure}
\hbox{\hspace{0.1cm}\epsfxsize=8.0cm \epsfbox{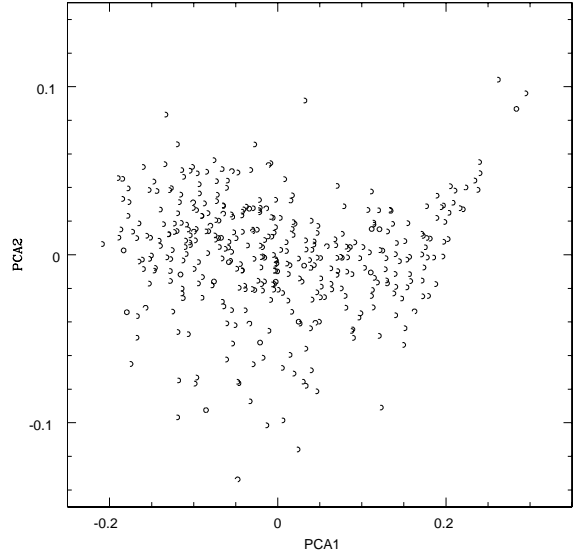}}
\caption{ Plot of first Principal Component against second Principal Component.}
\label{ref}
\end{figure}

\begin{figure}
\hbox{\hspace{0.1cm}\epsfxsize=8.0cm \epsfbox{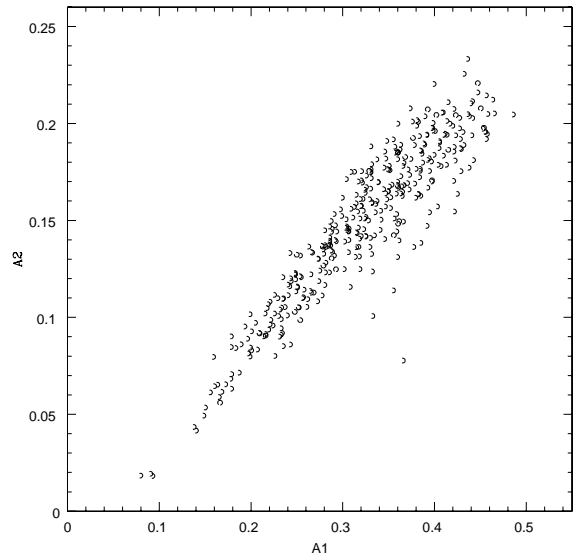}}
\caption{ Plot of first Fourier amplitude against second Fourier amplitude.}
\label{ref}
\end{figure}

\begin{figure}
\hbox{\hspace{0.1cm}\epsfxsize=8.0cm \epsfbox{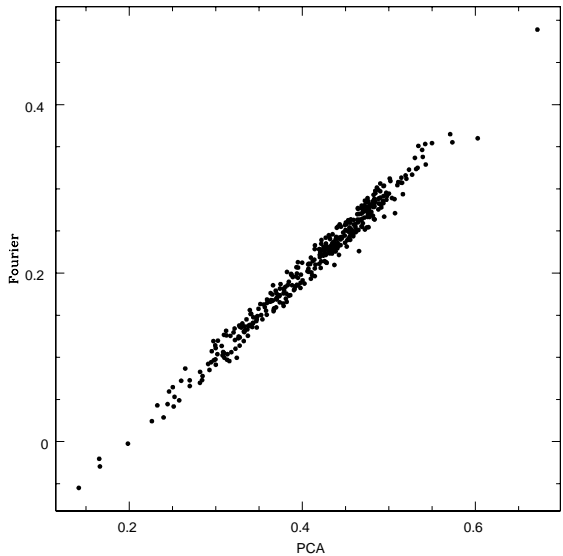}}
\caption{ Plot of fitted $M_v + const$ values when using Fourier and PCA methods.}
\label{ref}
\end{figure}

\begin{figure}
\hbox{\hspace{0.1cm}\epsfxsize=8.0cm \epsfbox{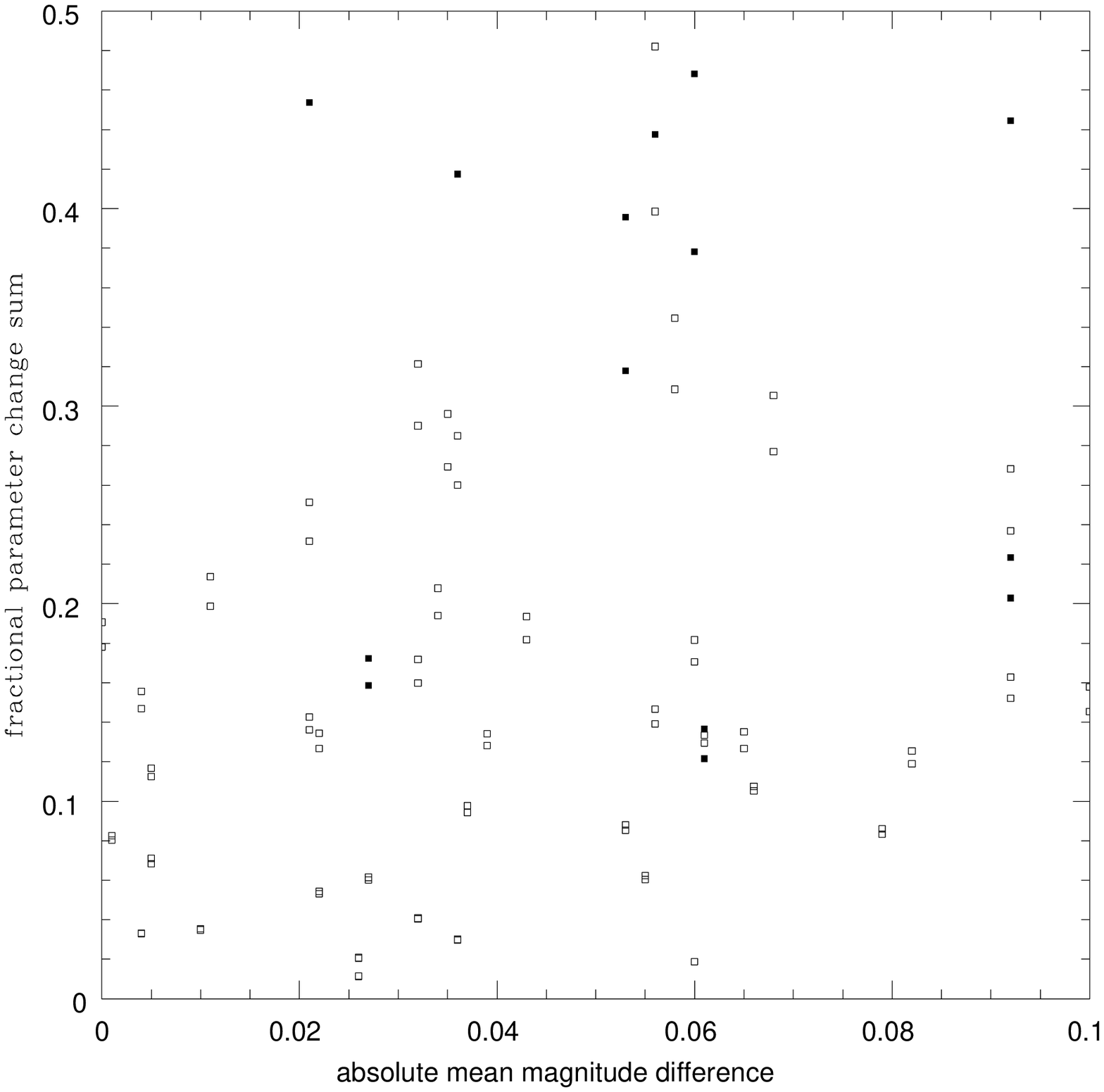}}
\caption{ Plot of absolute magnitude difference versus fractional change in light curve
parameters for Fourier (open squares) and PCA (closed squares).}
\label{ref}
\end{figure}

\section{Light Curve luminosity relations}

A major goal of stellar pulsation studies is to find formulae linking
global stellar parameters such as luminosity or metallicity to structural light
curve properties. If we are interested in the $V$ band magnitude, then we can write,
$$M_v=f(light curve structure),$$
where, since we do not know the function $f$, we try to estimate it
empirically. Two different approaches to quantifying light curve structure will,
in general, yield different formulations of the function $f$, but if there does exist
a true underlying function $f$, then both methods should give similar answers for $M_v$, given the
same input data. With a Fourier based method, the function $f$ is related to the
Fourier amplitudes and phases, $A_k, {\phi}_{k1}$, usually with a linear relation. With a PCA
approach, we use the PCA scores plotted in figures 2-4. Hence a PCA relation, though also linear,
will be different. The nature of PCA implies that the error structure in such formulae will be
simpler and we quantify this below. Both formulations should, of course, give similar
numbers for the final estimated value of the physical parameter
in question, in this case, $M_v$. 

KW used the Fourier method and found relations of the form,
\begin{eqnarray}
M_v = const. -  1.82\log P - 0.805A_1,
\end{eqnarray}
and,
\begin{eqnarray}
M_v = const. -1.876\log P -1.158A_1 + 0.821A_3.
\end{eqnarray}

We note that these relations were obtained through an iterative procedure
whereby
outliers were removed and the relations re-fitted (Kovacs 2004). In this paper, we use
the PCA method, but also, we use the entire dataset C mentioned in KW, consisting of 383
stars, and
fit the relations just once. We do not remove any outliers.
This may be why we obtain slightly different
versions of the fit using Fourier 
parameters than that published in KW. For ease of comparison, we include in
table 2 results obtained using both PCA and Fourier parameters. This table
gives the name for the relation, the independent variables considered
and coefficients together with their standard errors.
The value of
chi-squared in the table is defined as
\begin{eqnarray}
\sum_{k=1}^{k=N}(M_v - \hat{M_v})^2/(N-p),
\end{eqnarray}
where $\hat{M_v}$ is the fitted value of $M_v$ and $N,p$ are the number of
stars and parameters respectively in the fit. 
An examination of this table strongly suggests that
\begin{itemize}
\item{1)} Similar relations to equations (4) and (5) between $M_v$ and the PCA
coefficients exist.
\item{2)} We can use an F test (Weisberg 1980) to test for the significance
of adding a second and then a third PCA parameter to the regression. The F
statistic we use is
\begin{eqnarray}
{(RSS_{NH} - RSS_{AH})/(df_{NH}-df_{AH})}\over{RSS_{AH}/df_{AH}},\end{eqnarray}
where $RSS_{NH}, RSS_{AH}$ are the residual sum of squares under the
null and alternate (NH and AH) hypothesis respectively. Similarly,
$df_{NH}$ and $df_{AH}$ are the degrees of freedom under these two
hypotheses. For this problem, the null hypothesis is that the model
with the smaller number of parameters is sufficient whilst the
alternative hypothesis is that the model with the greater number of
parameters is required. Under the assumption of normality of errors,
equation (7) is distributed as an $F_{(df_{NH}-df_{AH}),df_{AH}}$,
(Weisberg, 1980, p. 88).
Applying this $F$ test implies firstly, that adding the first parameter
$PCA1$ is a significant addition to $\log P$ and secondly, that 
adding a second and third parameter, $PCA2$ and $PCA3$ are also highly
significant with a p value less than 0.0004.
In the case of Fourier parameters, adding the $A_1$ parameter to $\log P$
is highly significant and adding the $A_3$ parameter to this is also
highly significant. However, a formula involving ($\log P, A_1, A_2$)
has a p value of 0.0058 and a formula involving all 3 Fourier amplitudes
and $\log P$ is not a significant addition to a formula involving
$(\log P, A1, A3)$.
\item{3)} The standard deviation of the fits given in the last
column is generally slightly higher for the PCA case, when considering similar
numbers of parameters. This is perhaps caused by the fact that the different
PCA components carry orthogonal sets of information.
\item{4)} The errors on the coefficients in the PCA fits are
always significantly smaller. This is an important point when we evaluate the
errors on the final fitted value of the absolute magnitude.
\item{5)} If we write the absolute magnitude as a function of
parameters, $x_1, x_2,..,x_N$,
\begin{eqnarray}
M_v + const. = f(x_1,x_2,...,x_N),\end{eqnarray}
then the error on the absolute magnitude is given by,
\begin{eqnarray}
{\sigma}^2(M_V + const.) = \sum_{k=1}^{k=N}{\sigma}^2(x_k){( {\partial f \over
{\partial x_k}})}^2 +\nonumber\\ \sum_{i,j=1,i\ne j}^{N}{\sigma}^2(x_i,x_j){({\partial f\over{\partial x_i}})} {({\partial f\over{\partial x_j}})} .
\end{eqnarray}
As table 2 indicates, ${\sigma}^2(x_k)$ is always smaller when the
$x_k$ are PCA coefficients rather than Fourier amplitudes. Figure 8 and 9
portray graphs of $PCA1$ vs $PCA2$ and $A_1$ verses $A_2$ respectively.
We note that ${\rho}_{i,j}{\sigma}(x_i){\sigma}(x_j) = {\sigma}^2(x_i,x_j)$.
Table 3 presents sample correlation and covariance coefficients between the
period and PCA
parameters and period and Fourier parameters. Table 3, and  figures 6 and 7
demonstrate that the correlation coefficient
amongst any pair of PCA coefficients is smaller than between any pair of
Fourier coefficients. Hence the error on the fitted value of
$M_v$, ${\sigma}^2(M_v)$, {\it has} to be smaller when using a
PCA based formula. We can use table 3 and equation (9) to formally
calculate the error on $M_v+const$.
Table 4 presents these results. The label in the top row of this table (P1, F1, etc.,) refers to
the appropriate relation in table 2. We see clearly
that the PCA formulae do better than their Fourier counterparts
with a similar number of parameters. When we consider the
$(\log P, PC1, PC2)$ and $(\log P, A_1, A_3)$ variables,
then the "error advantage" using a PCA based method is a factor of two. 
This occurs not just because the PCA coefficients 
are orthogonal to each other, but also because the errors on the 
coefficients in a PCA based formula are significantly smaller
than in the Fourier case.

\end{itemize}

Figure 10 displays a plot of the predicted absolute magnitudes obtained
using a two parameter ($\log P, A_1, A_3$) Fourier fit and the three parameter ($\log P, PCA1,
PCA2, PCA3$) PCA fit.
The two approaches are displaced from each other because we do not consider
the constants in this study. Disregarding this, it can be seen that the slope
of this plot is 1: hence the two methods produce
similar relative absolute magnitudes.

\section{Conclusion}

We have shown that the method of PCA can be used to study RR Lyrae light curves.
It has distinct advantages over a Fourier approach because
\begin{itemize}
\item{a)} It is a more efficient way to characterize structure since
fewer parameters are needed. A typical Fourier fit requires 17 parameters
whereas a PCA fit may only need 9.
\item{b)} Using the PCA approach, we see clearly why the amplitude is a good
descriptor
of RRab light curve shape.
\item{c)} The different PCA components are orthogonal to each other whereas
the Fourier amplitudes are highly correlated with each other. This leads to
relations linking light curve structure to absolute magnitude using PCA
having coefficients with smaller errors and leading to more accurate
estimates of absolute magnitudes. This can reduce the formal error,
in some cases, by a factor of 2.
\end{itemize}

In the present formulation of our PCA approach, the input data is a Fourier analysis. If these
input data, that is the Fourier decompositions, contain significant observational errors, the error
bars on the resulting Principal Components will be larger. Neither the PCA or Fourier approach can
compensate fully for noisy data. In this sense, the sensitivity of PCA to noisy data should be similar to
Fourier, though the fact that PCA is an ensemble approach in which we initially remove an average term does
guard against individual points having too much undue influence. As an example, table 4 of KW gives 17 outliers
(in
terms of their Fourier parameters),
which KW removed in their analysis relating absolute magnitude to Fourier parameters.
We do {\it not} remove these
outliers, yet, in terms of the final fitted magnitudes presented in figure 10, PCA and Fourier produce very
similar results. Further, even with the inclusion of these 17 stars, the
PCA method still produces PCA coefficients with smaller errors as given in tables 2 and 3. 
Kanbur et al (2002) discuss 
in detail the nice error properties of the PCA method as applied to variable stars and give a recipe with which to
calculate errors on PCA coefficients. Their figure 2, albeit for Cepheids, displays error bars on these coefficients. We see that even with noisy data,
the progression of PCA parameters with period is preserved, though of course, the error bars on the PCA
coefficients are larger.

Ngeow et al (2003) developed a simulated annealing method which can reduce numerical wiggles in Fourier
decomposition of sparse data. Ngeow et al (2003) give specific examples of how
such an approach improves Fourier techinues using OGLE LMC Cepheids. A similar result will hold true
for RR Lyraes. Hence this annealing technique couple with a Principal Component analysis should prove very
useful when dealing with noisy RR Lyrae data and will be treated in detail in a subsequent paper.

Our PCA results are based on a sample of 383 stars in globular clusters. How transferable are our results and how can our
results be used to obtain PC coefficients for a new RR Lyrae light curve which appears to be
normal (ie no signs of Blazhko effects etc.)?

Our results are transferable to the extent that the original 383 stars are a good
representation of the entire population of RRab stars, including variation in metallicity and
differences between field and cluster variables. Given this caveat, we suggest two methods to
reproduce the light curve of a new RRab star. Firstly, it is straightforward to include the new star in the
PCA analysis with the existing dataset. This is our recommended approach and preserves the "ensemble analysis"
property of our PCA method. Our second method will be the subject of future paper
but briefly it is this. We fit the progression of the PCA coefficients with
period, such as given in figures 4 and 5, with simple polynomial functions. As an aside, we
remark that figure 4 contains significant scatter, perhaps associated with
metallicity, so that it would be best to include metallicity in such polynomial fits.
For a new star, we then guess its period and read off, for that period, the value of the
PCA coefficients. Equation (3) then allows us to generate the light curve. We iterate this until a specified
error criterion is satisfied. We can then use existing formulae relating absolte magnitude to
light curve structure as defined by PCA. This PCA template approach has been used, with considerable success,
in analysing HST Cepheid data (Leonard et al 2003).

We note from table 2 that the chi square on the fitted relations are similar for PCA and Fourier.
Does this mean that despite the smaller formal errors with PCA, both methods' ability to predict
RRab absolte magnitudes is limited by the intrinsic properties of RRab stars themselves? To some
extent this is true. Jurcsik et al (2004), in analysing accurate data for 100
RRab stars in M3, show that for some 16
stars, amongst which there exist some pairs whose absolute mean magnitudes differ by about 0.05 mags (the accuracy of the photometry is about
0.02mags), the Fourier parameters and periods are very similar. That is, an empirical method relating absolute magnitude to
period and Fourier parameters in one waveband could not distinguish between these stars. Since, as Jurcsik et al (2004)
point out, their data contains a small range of both mass and metallicity, temperature is the only other variable, it may be the case
that multiwavelength information is needed. It is worthwhile to investigate how PCA fares with this
dataset. Here we give an outline that
suggests that PCA can be more efficient at extracting information from the light curve.

For the sixteen stars which had differing absolute magnitudes but very similar Fourier parameters, we can perform the
following procedure: for every pair, $j\ne k$, we calculate
$$(a1(j)-a1(k))/a1(k) + (a2(j)-a2(k))/a2(k)$$
$$ + (a3(j) - a3(k))/a3(k)=diff1,$$
$$(pca1(j)-pca1(k))/pca1(k) +$$
$$ (pca1(j)-pca2(k))/pca2(k)=diff2,$$
and
$$(vmean(j)-vmean(k))=diff3,$$
where $a1(j),a2(j),a3(j)$ are the Fourier amplitudes and $pca1(j),pca2(j)$
are the PCA coefficients and $vmean(j)$ are the mean magnitudes.
In the above, we always take the absolute value of the differences. We need to take fractional 
changes because the Fourier amplitudes and PCA coefficients have different ranges. We now plot diff3 against diff1
and diff2. This is presented in figure 11, where the open squares are diff1 and the closed squares are diff2.
We see that with PCA (closed squares), the differences between light curve structure parameters are greater than with Fourier (open squares).
This could imply that PCA can be more efficient though the limitations associated with using a single
waveband are still present. 
A more rigorous, quantitative discussion of this, in a Fisher information sense, will be given in a future
paper.

In other future work we plan to investigate the applicability of this method to light curve structure-metallicity
relations, RRc stars and a comparison of observed and theoretical light curves
using PCA.

\begin{table}
\centering
\caption{Percentage of variation explained by PC components}
\label{tab1}
\begin{tabular}{lcccccc} \\ \hline
&PC1&PC2&PC3&PC4&PC5&PC6 \\ \hline
without average&81.4&7.8&5.7&2.3&0.74&0.57 \\
with average&96.9&1.9&0.55&0.25&0.07&0.006\\
\hline
\end{tabular}
\end{table}

\begin{table*}
\centering
\caption{Light curve luminosity relation using PCA and Fourier methods.}
\label{tab1}
\begin{tabular}{lcccccc} \\ \hline
&$\log P$&first&second&third&chisquare \\ \hline
PCA \\ \hline
P0&-1.134$\pm0.059$&&&&0.00321 \\
P1&-1.550$\pm0.082$&0.269$\pm0.038$&&&0.00283 \\
P2&-1.609$\pm0.082$&0.290$\pm0.038$&0.291$\pm0.082$&&0.00274 \\
P3&-1.744$\pm0.088$&0.329$\pm0.039$&&-0.539$\pm0.107$&0.0027 \\
P4&-1.829$\pm0.088$&0.359$\pm0.039$&0.336$\pm0.079$&-0.583$\pm0.105$&0.00253 \\ \hline
Fourier \\ \hline
F1&-1.677$\pm0.083$&-0.472$\pm0.054$&&&0.00266 \\
F2&-1.700$\pm0.082$&-0.726$\pm0.092$&&0.613$\pm0.179$&0.00258 \\
F3&-1.740$\pm0.085$&-0.758$\pm0.116$&0.536$\pm0.193$&&0.00261 \\
F4&-1.720$\pm0.085$&-0.790$\pm0.117$&0.215$\pm0.243$&0.490$\pm0.227$&0.00258 \\
\hline
\end{tabular}
\end{table*}

\begin{table*}
\centering
\caption{Sample correlation and covariance coefficients between period, PCA
and Fourier coefficients}
\label{tab2}
\begin{tabular}{lccccccc}\\ \hline
&$\log P, PCA1$&$\log P, PCA2$&$\log P, PCA3$&$PCA1,PCA2$&$PCA2,PCA3$&$PCA1,PCA3$& \\ \hline
correlation&0.631&0.099&-0.299&$<10^{-6}$&$<10^{-6}$&$<10^{-6}$& \\
covariance&0.0038&0.0002&-0.0006&$<10^{-6}$&$<10^{-6}$&$<10^{-6}$&\\
\hline
&$\log P, A_1$&$\log P, A_2$&$\log P, A_3$&$A_1, A_2$&$A_2, A_3$&$A_1, A_3$&\\
\hline
correlation&-0.655&-0.529&-0.562&0.926&0.931&0.902&\\
covariance&-0.0028&-0.0012&-0.0011&0.0030&0.0013&0.0024&\\

\end{tabular}
\end{table*}

\begin{table*}
\centering
\caption{Formal error on $M_v+const.$ for PCA and Fourier relations}
\label{tab3}
\begin{tabular}{lcccccccc}\\ \hline
P1&P2&P3&P4&F1&F2&F3&F4 \\
0.0139&0.0142&0.0216&0.0240&0.0156&0.0313&0.0394&0.0311\\ \hline
\end{tabular}
\end{table*}

\section*{acknowledgments}

SMK thanks Geza Kovacs for stimulating discussions and for kindly supplying the
RRab dataset. SMK thanks D. Iono for help writing the PCA and least squares program and C. Ngeow
for help with latex. HM thanks FCRAO for providing a summer internship in 2001 when part of this work
was completed. We also thank the referee for constructive comments.




\begin{thebibliography}{}
\bibitem[1999]{hen99} Hendry, M. A., Tanvir, N. A., Kanbur, S. M., 1999, ASP Conf. Series, 167, p. 192
\bibitem[2003]{kan03} Kanbur, S., Iono, D., Tanvir, N. \& Hendry, M., 2002, MNRAS, 329, 126
\bibitem[2004]{jur04} Jurcsik, J., Benko, J., Bakos, G., Szeidl, B., Szabo, R., 2004, ApJL, 597, 49
\bibitem[2001]{kov01} Kovacs, G., Walker, A. R., 2001, A\&A, 371, 579
\bibitem[2004]{kov04} Kovacs, G., 2004, private communication
\bibitem[2003]{leo03} Leonard, D., Kanbur, S.,M., Ngeow, C., Tanvir, N., 2003, ApJ, 594, 247
\bibitem[2003]{ngeo03} Ngeow, C., Kanbur, S. M., Nikolaev, S., Tanvir, N., Hendry M., 2003, 586, 959
\bibitem[2004]{tan04} Tanvir, N. R., Hendry, M. A., Kanbur, S. M., 2004, in preparation
\bibitem[1980]{wei80} Weisberg, S., 1980, {\it Applied Linear Regression}, John Wiley \& Sons, $1^{st}$ Ed.
\end{thebibliography}
\end{document}